\documentclass[twocolumn,showpacs,preprintnumbers,amsmath,amssymb,
    floatfix,prl]{revtex4}

\usepackage{graphicx}
\usepackage{epsfig} 
\usepackage{dcolumn}
\usepackage{bm}
\usepackage{subfigure} 

\newcommand{\order}[1]{\mathcal{O}\left(#1\right)}
\newcommand{\avg}[1]{\left\langle#1\right\rangle}
\newcommand{\paren}[1]{\left(#1\right)}
\newcommand{\bracket}[1]{\left\lbrack#1\right\rbrack}
\newcommand{\Deltabar}{\overline{\Delta}}

\begin{document}


\title{Chimera States for Coupled Oscillators}

\author{Daniel M. Abrams}
  \email{dma32@cornell.edu}
\author{Steven H. Strogatz}
  \email{shs7@cornell.edu}
\affiliation{
212 Kimball Hall, Department of Theoretical and Applied Mechanics,\\
Cornell University, Ithaca, NY 14853-1503, USA }


\begin{abstract}
Arrays of identical oscillators can display a remarkable spatiotemporal
pattern in which phase-locked oscillators coexist with drifting ones.
Discovered two years ago, such ``chimera states" are believed to be
impossible for locally or globally coupled systems; they are peculiar to the
intermediate case of nonlocal coupling. Here we present an exact solution for
this state, for a ring of phase oscillators coupled by a cosine kernel.  We
show that the stable chimera state bifurcates from a spatially modulated
drift state, and dies in a saddle-node bifurcation with an unstable chimera.
\end{abstract}



\maketitle

In Greek mythology, the chimera was a fire-breathing monster having a lion's
head, a goat's body, and a serpent's tail.  Today the word refers to anything
composed of incongruous parts, or anything that seems fantastical.

This paper is about a mathematical chimera in which an array of identical
oscillators splits into two domains: one coherent and phase-locked, the other
incoherent and desynchronized \cite{kur02, kur03, shima04}. Nothing like this
has ever been seen for identical oscillators.  It cannot be ascribed to a
supercritical instability of the spatially uniform oscillation, because it
occurs even if the uniform state is stable. Furthermore, it has nothing to do
with the partially locked/partially incoherent states seen in populations of
non-identical oscillators with distributed frequencies \cite{winfree67,
kur84}. There, the splitting of the population stems from the inhomogeneity
of the oscillators themselves; the desynchronized oscillators are the
intrinsically fastest or slowest ones. Here, all the oscillators are the
same.

In this Letter we explain where the chimera comes from and pinpoint the
conditions that allow it to exist.  It was first noticed by Kuramoto and his
colleagues \cite{kur02, kur03, shima04} while simulating arrays of
limit-cycle oscillators with nonlocal coupling. As they emphasize, nonlocal
coupling \cite{ermentrout85, barahona02, bressloff98} is less explored than
local or global coupling \cite{winfree67, kur84, strogatz00, peles03,
bonilla98, daido96}. It arises in diverse applications ranging from Josephson
junction arrays \cite{phillips93} and chemical oscillators \cite{kur02,
kur03, shima04}, to the neural networks underlying snail shell patterns
\cite{murraybook, ECO} and ocular dominance stripes \cite{murraybook,
swindale80}.

We study the simplest system that supports a chimera state: a ring of phase
oscillators \cite{kur02, kur03} governed by
\begin{equation}\label{maineq}
   \frac{\partial \phi}{\partial t} = \omega -
   \int_{-\pi}^{\pi}{G\paren{x-x'}\sin \bracket{\phi(x,t) - \phi(x',t) + \alpha} dx'}~.
\end{equation}
Here $\phi(x,t)$ is the phase of the oscillator at position $x$ at time $t$.
The space variable $x$ runs from $-\pi$ to $\pi$ with periodic boundary
conditions. The frequency $\omega$ plays no role in the dynamics; one can set
$\omega = 0$ by redefining $\phi \rightarrow \phi + \omega t$ without
otherwise changing the form of Eq.~(\ref{maineq}). The angle $0 \le \alpha
\le \frac{\pi}{2}$ is a tunable parameter. The kernel $G(x-x')$ provides
nonlocal coupling between the oscillators.  It is assumed to be even,
non-negative, decreasing with the separation $|x-x'|$ along the ring, and
normalized to have unit integral.  Kuramoto and Battogtokh \cite{kur02,
kur03} assumed an exponential kernel $G(x) \propto \exp(-\kappa|x|)$, but
instead we will take
\begin{equation}\label{kerneldef}
   G(x) = \frac{1}{2\pi} (1+A\cos x)
\end{equation}
where $0 \le A \le 1$.  Simulations show that both kernels give qualitatively
similar results, but the cosine kernel allows the model to be solved
analytically.

\begin{figure}[t] 
   \centerline{
     \epsfig{file=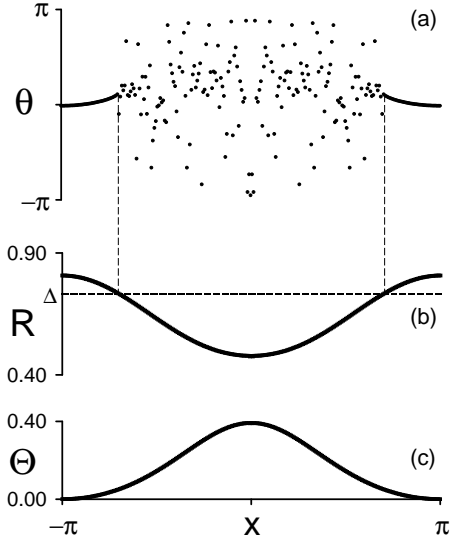, scale=0.35}
   }
   \caption{
     (a) Phase pattern for a chimera state. Parameters: $A = 0.995$,
     $\beta = 0.18$, $N = 256$ oscillators. Equation \eqref{maineq} was
     integrated using a Runge-Kutta method with fixed time step $dt = 0.025$
     for 200,000 iterations, starting from
     $\phi(x) = a \exp{\bracket{-b x^2}} r$,
     where $r$ is a uniform random variable on
     $[-\frac{1}{2}, \frac{1}{2}]$.  We took $a = 6$ and $b=30$.
     (b) Local phase coherence $R(x)$, computed from \eqref{orderparam}.
     Locked oscillators satisfy $R(x) \ge \Delta$.
     (c) Local average phase $\Theta(x)$.}
   \label{fig:numint}
\end{figure}

Figure~\ref{fig:numint}(a) shows a snapshot of a chimera state for
Eq.~\eqref{maineq}. The oscillators near $x=\pm\pi$ are locked and coherent:
they all move with the same instantaneous frequency and are nearly in phase.
Meanwhile, the scattered oscillators in the middle of Figure 1(a) are
drifting, both relative to each other and relative to the locked oscillators.
They slow down as they pass the locked pack, which is why the dots appear
more densely clumped there.

These simulation results can be explained \cite{kur02} by generalizing
Kuramoto's earlier self-consistency argument for globally coupled oscillators
\cite{kur84, strogatz00}. Let $\Omega$ denote the angular frequency of a
rotating frame in which the dynamics simplify as much as possible, and let
$\theta = \phi - \Omega t$ denote the phase of an oscillator relative to this
frame.  Introduce a complex order parameter $R e^{i \Theta}$ that depends on
space and time:
\begin{equation}\label{orderparam}
   R(x,t) e^{i \Theta(x,t)} = \int_{-\pi}^{\pi}{G\paren{x-x'} e^{i
      \theta(x',t)} dx' }~.
\end{equation}
Then Eq.(\ref{maineq}) becomes
\begin{equation}\label{decoupled}
   \frac{\partial \theta}{\partial t} = \omega - \Omega - R
   \sin \bracket{\theta - \Theta + \alpha}~.
\end{equation}
By restricting attention to stationary solutions, in which $R$ and $\Theta$
depend on space but not on time (a condition that also determines $\Omega$),
Kuramoto and Battogtokh \cite{kur02} derived a self-consistency equation
equivalent to
\begin{eqnarray}\label{SCnewvars}
   R(x) \exp{[i \Theta(x)]} &=& e^{i \beta} \int_{-\pi}^{\pi}{G(x-x')
\exp{[i \Theta(x')]}} \nonumber \\
   & & {} \times {\frac{\Delta - \sqrt{\Delta^2 - R^2(x')}}{R(x')}} dx'
\end{eqnarray}
where $ \beta = \frac{\pi}{2} - \alpha$ and $\Delta = \omega -
\Omega$.

Equation (\ref{SCnewvars}) is to be solved for three unknowns---the
real-valued functions $R(x)$ and $\Theta(x)$ and the real number
$\Delta$---in terms of the assumed choices of $\beta$ and the kernel $G(x)$.
Kuramoto and Battogtokh \cite{kur02,kur03} solved (\ref{SCnewvars})
numerically via an iterative scheme in function space, and confirmed that the
resulting graphs for $R(x)$ and $\Theta(x)$ match those obtained from
simulations of Eq.(\ref{maineq}).  Figures~\ref{fig:numint}(b)
and~\ref{fig:numint}(c) show the graphs of $R(x)$ and $\Theta(x)$ for the
parameters used in Figure~\ref{fig:numint}(a).

The resemblance of these curves to cosine waves suggested to us that
Eq.(\ref{SCnewvars}) might have a closed-form solution. It does.  Since the
right hand side of (\ref{SCnewvars}) is a convolution integral, the equation
is solvable for any kernel in the form of a finite Fourier series; that is
what motivated the choice of (\ref{kerneldef}). For this case, $R(x)$ and
$\Theta(x)$ can be obtained explicitly. The resulting expressions, however,
still contain two unknown coefficients, one real and the other complex, that
need to be determined self-consistently.

The solution proceeds as follows.  Let
\begin{equation}\label{hdef}
   h(x') = \frac{\Delta-\sqrt{\Delta^2-R^2(x')}}{R(x')}
\end{equation}
and let angular brackets denote a spatial average:
\begin{displaymath}
   \langle f \rangle = \frac{1}{2\pi} \int_{-\pi}^{\pi}{f(x')dx'}~.
\end{displaymath}
Expanding $G$ with a trigonometric identity, Eq.(\ref{SCnewvars}) gives
\begin{eqnarray}\label{newSC}
     R e^{i \Theta} &=& e^{i \beta} \langle {h e^{i \Theta}} \rangle +
A e^{i \beta} \langle {h e^{i \Theta} \cos x'} \rangle \cos x
\nonumber \\
     & & \qquad\qquad\qquad {} + A e^{i \beta} \langle {h e^{i \Theta}
\sin x'} \rangle \sin x \nonumber \\
     &=& c + a \cos x
\end{eqnarray}
where the unknown coefficients $a$ and $c$ are given by
\begin{eqnarray}
   \label{cdef} c &=& e^{i \beta} \langle {h e^{i \Theta}} \rangle \\
   \label{adef} a &=& A e^{i \beta} \langle {h e^{i \Theta} \cos x'} \rangle~.
\end{eqnarray}
The coefficient of $\sin x$ vanishes in (\ref{newSC}), if we assume $R(x') =
R(-x')$ and $\Theta(x') = \Theta(-x')$, as suggested by the simulations. The
assumed evenness is self-consistent: it implies formulas for $R(x)$ and
$\Theta(x)$ that indeed possess this symmetry. For instance, $R(x)$ satisfies
\begin{eqnarray}\label{Rsquared}
     R^2 &=& (R e^{i \Theta}) (R e^{-i \Theta}) \nonumber \\
     &=& (c + a \cos x) (c^* + a^* \cos x) \nonumber \\
     &=& |c^2| + 2 \textrm{Re} (c a^*) \cos x + |a|^2 \cos^2 x~,
\end{eqnarray}
which also explains why the graph in Fig.~\ref{fig:numint}(b) resembles a
cosine.  Likewise, $\Theta(x)$ satisfies
\begin{equation}\label{tanTheta}
     \tan \Theta(x) = \frac{R(x) \sin \Theta(x)} {R(x) \cos \Theta(x)}
                    = \frac{c_i + a_i \cos x} {c_r + a_r \cos x}
\end{equation}
where the subscripts denote real and imaginary parts.

Another simplification is that $c$ can be taken to be purely real and
non-negative, because of the rotational symmetry of the governing equations.
In particular, the self-consistency equation (\ref{SCnewvars}) is left
unchanged by any rigid rotation $\Theta(x) \rightarrow \Theta(x) + \Theta_0$.
Thus we are free to specify any value of $\Theta(x)$ at whatever point $x$ we
like. We choose $\Theta(\frac{\pi}{2}) = 0$. Then $R e^{i \Theta} = c + a
\cos x$ implies $R(\frac{\pi}{2}) = c$. Since $R$ is real and non-negative,
so is $c$. Hence, we take $c_i = 0$ from now on.

To close the equations for $a$ and $c$, we rewrite the averages in
(\ref{cdef}) and (\ref{adef}) in terms of those variables. Using
\begin{displaymath}
  h e^{i \Theta} = \left(R e^{i \Theta}\right) \frac{h}{R}
                 = \frac{\Delta - \sqrt{\Delta^2 - R^2(x)}}{c + a^* \cos x}
\end{displaymath}
and inserting (\ref{Rsquared}) into (\ref{cdef}) and (\ref{adef}), we find
%
\begin{equation}\label{ceqn}
   c = e^{i \beta} \avg{ {\frac{\Delta-\paren{\Delta^2 - c^2 - 2 c a_r \cos x
       - |a|^2\cos^2 x}^{\frac{1}{2}}} {c + a^* \cos x}} }
\end{equation}
%
\begin{equation} \label{aeqn} \textstyle
   a = A e^{i \beta} \avg{ { \frac{\Delta-\paren{\Delta^2
       - c^2 - 2 c a_r \cos x - |a|^2\cos^2 x}^{\frac{1}{2}}} {c + a^* \cos x}
       \cos x} }~.
\end{equation}
This pair of complex equations is equivalent to four real equations for the
four real unknowns $c$, $a_r$, $a_i$, and $\Delta$.  The solutions, if they
exist, are to be expressed as functions of the parameters $\beta$ and $A$.

\begin{figure}[t] 
   \centerline{
     \epsfig{file=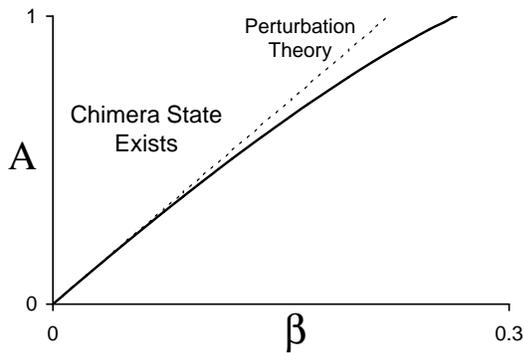, scale=0.32}
   }
   \caption{Region in parameter space where a chimera state
exists. Solid line, boundary determined by numerical solution of
Eqs.~\eqref{ceqn}, \eqref{aeqn}; dashed line, approximate boundary obtained
from perturbation theory (see text). A saddle-node bifurcation of a stable
and unstable chimera occurs at the boundary.}
   \label{fig:wedge}
\end{figure}

Figure~\ref{fig:wedge} plots the region in parameter space where chimera
states exist, computed by solving Eqs.~\eqref{ceqn}, \eqref{aeqn}, with a
root-finder and numerical continuation. To obtain starting guesses for the
four unknowns, we integrated \eqref{maineq} numerically, then fit the
resulting $R(x)$ and $\Theta(x)$ to the exact solutions \eqref{Rsquared},
\eqref{tanTheta}, to extract the corresponding $c$, $a_r$, and $a_i$, and
estimated $\Delta$ directly from the collective frequency of the locked
oscillators.  By sweeping $\beta$ at fixed $A$, we found that the chimera
state disappeared suddenly when it reached the boundary of the region.

For deeper insight into the chimera state and its bifurcations, we now solve
Eqs. \eqref{ceqn}, \eqref{aeqn} perturbatively.  Figure \ref{fig:wedge}
suggests that we should allow $\beta$ and $A$ to tend to zero simultaneously:
let $A=\epsilon$, $\beta = \beta_1 \epsilon$, and seek solutions of
\eqref{ceqn}, \eqref{aeqn} as $\epsilon \rightarrow 0$.  Numerical
continuation reveals that the solutions behave as follows:
\begin{eqnarray}\label{ansatz}
   \Delta & \sim & 1 + \Delta_1 \epsilon + \Delta_2 \epsilon^2 \nonumber \\
   c & \sim & 1 + c_1 \epsilon + c_2 \epsilon^2 \nonumber \\
   a_r & \sim & a_{2r} \epsilon^2 \nonumber \\
   a_i & \sim & a_{2i} \epsilon^2
\end{eqnarray}
where terms of $\order{\epsilon^3}$ have been neglected. Substituting this
ansatz into \eqref{ceqn}, \eqref{aeqn}, we find that $\Delta_1 = c_1$ is
required to match terms of $\order{\sqrt{\epsilon}}$. Then at leading order,
the real and imaginary parts of Eqs.~\eqref{ceqn}, \eqref{aeqn} become
\begin{eqnarray}
  \label{c1eqn}
  c_1     &=& -\textrm{Re}{\bracket{\sqrt{2} \avg{\sqrt{\delta - u \cos x}}}} \\
  \label{betaeqn}
  \beta_1 &=&  \textrm{Im}{\bracket{\sqrt{2} \avg{\sqrt{\delta - u \cos x}}}} \\
  \label{ueqn}
  u      &=& -\textrm{Re}{\bracket{\sqrt{2} \avg{\cos x \sqrt{\delta - u \cos x}}}}\\
  \label{a2ieqn}
  a_{2i}  &=& -\textrm{Im}{\bracket{\sqrt{2} \avg{\cos x \sqrt{\delta - u\cos x}}}}
\end{eqnarray}
where we've defined $\delta = \Delta_2 - c_2$ and $u=a_{2r}$.

The solutions of these equations can be parametrized by $\delta$, as follows.
Writing $f(u,\delta)$ for the right hand side of Eq.~\eqref{ueqn}, we compute
all the real roots of $u = f(u,\delta)$, and regard them as functions of
$\delta$.  Then we sweep through all the $\delta$'s for which
Eq.~\eqref{ueqn} has a solution, and substitute the associated $u(\delta)$
into the remaining equations to generate values of $c_1(\delta)$,
$\beta_1(\delta)$, and $a_{2i}(\delta)$.

\begin{figure}[t] 
   \centerline{
     \epsfig{file=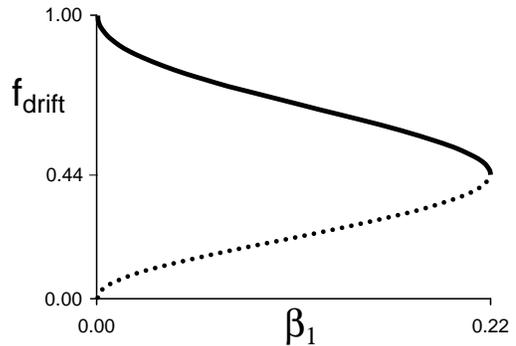, scale=0.3}
   }
   \caption{Fraction of oscillators in the chimera state that drift.
   Solid line, stable chimera; dotted line, unstable chimera.}
   \label{fig:fdrift}
\end{figure}

This approach gives a great deal of information about the chimera state. For
instance, Figure \ref{fig:fdrift} plots the fraction of oscillators that are
drifting, $f_{\textrm{drift}} = \frac{1}{\pi} \cos^{-1}(\delta/u(\delta))$,
as a function of the control parameter $\beta_1(\delta)$.  There are two
branches of solutions.  The upper branch (which dynamical simulations of
Eq.~\eqref{maineq} show to be stable) bifurcates from a state of pure drift
at $\beta_1 = 0$. As $\beta_1$ increases, drifting oscillators are
progressively converted into locked ones, eventually reaching a minimum of
about 44\% drift at the largest $\beta_1$ for which stable chimeras exist,
$\paren{\beta_1}_{\textrm{max}} \approx 0.2205$.  There the upper branch
collides with the lower (unstable) one, which itself emerges from a
homoclinic locked state at $\beta_1 = 0$, where the in-phase oscillation of
Eq.~\eqref{maineq} is linearly neutrally stable. The maximum value of
$\beta_1$ predicts that the slope of the stability boundary in
Fig.~\ref{fig:wedge} equals $1 / \paren{\beta_1}_{\textrm{max}} \approx
4.535$, shown there as a dashed line.

Unfortunately, when the variables are plotted versus $\beta_1$, much of the
bifurcation structure is hidden.  In particular, two crucial events in the
genesis of the chimera state occur when $f_{\textrm{drift}} = 1$, $\beta_1 =
0$, and therefore collapse onto a single point in Figure \ref{fig:fdrift}.
These events create the $x$-dependence in the chimera state, first in its
local coherence $R(x)$ and then in its local average phase $\Theta(x)$.  To
see how such spatial structure arises, it is best to treat $\delta$, not
$\beta_1$, as the relevant parameter, even though it is not a true control
parameter (its turning points do not signify bifurcations, for example).

\begin{figure}[t] 
   \centerline{
     \epsfig{file=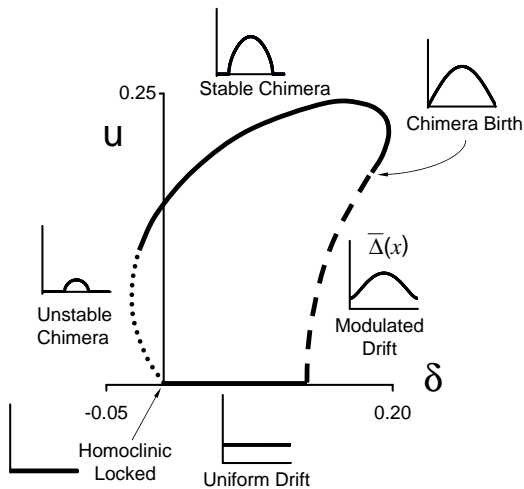, scale=0.4}
   }
   \caption{Solutions of Eq.~\eqref{ueqn} in $(\delta, u)$ plane, where
   $\delta = \Delta_2 - c_2$ and $u = a_{2r}$. Inset panels show typical
   mean drift frequencies $\Deltabar(x)$ in different regions of solution
   space.  Panels with arrows indicate the shape of $\Deltabar(x)$ for
   transitional values of $(\delta,u)$.}
   \label{fig:u-delta}
\end{figure}

Figure~\ref{fig:u-delta} plots $u(\delta)$ vs.~$\delta$ for the roots of
Eq.~\eqref{ueqn}. Branches have been coded with different dashing styles to
indicate that they represent qualitatively different states.  The zero branch
along the $\delta$-axis, shown as a solid line, represents a family of
spatially uniform drift states where both $R$ and $\Theta$ are independent of
$x$. Such states occur only when $\beta = 0$. They correspond to the exact
(non-perturbative) solutions of Eq.~\eqref{SCnewvars} given by $\Theta(x)
\equiv 0$ and $R(x) \equiv R = \sqrt{2 \Delta - 1}$, with $\frac{1}{2} \le
\Delta < 1$. Linearization shows that this uniform drift state undergoes a
zero-eigenvalue bifurcation at $\Delta = \frac{1}{2} (2+\epsilon)^{-2}$,
valid for all $0 < \epsilon < 1$, implying a critical value of $\delta =
\frac{1}{8}$ as $\epsilon \rightarrow 0$.

This is the first crucial event.  At $\delta = \frac{1}{8}$, a spatially
modulated drift state is born. Now $R$ depends on $x$.  In perturbative
variables, a root with $u = a_{2r} \neq 0$ bifurcates off the zero branch
(shown dashed in Figure~\ref{fig:u-delta}). Meanwhile, $a_{2i}=0$, so we
still have $\Theta(x) \equiv 0$ for all $x$ (from Eq.~\eqref{tanTheta} and
$c_i = 0$). Thus for states on the dashed branch, all the oscillators are
drifting, and maintain the same average phase, but with different amounts of
coherence at different values of $x$. Like the uniform drift states, these
modulated drift states occur only if $\beta = 0$.

The second crucial event occurs when the dashed branch intersects the line $u
= \delta$. Then several things happen.  The first locked oscillators are
born; $a_{2i}$ and $\beta_1$ become nonzero; $\Theta$ depends on $x$; and a
stable chimera is created. Evaluating the integral in \eqref{ueqn} for $u =
\delta$ shows that $\delta = 16/9\pi^2 \approx 0.18$ at the birth of the
chimera.

Another way to distinguish among the various states is shown in the insets of
Figure~\ref{fig:u-delta}. For selected values of $\delta$ we have plotted the
time-averaged frequency $\Deltabar(x)$ of the oscillator at $x$, measured
relative to the rotating frame.  For locked oscillators, $\Deltabar(x) = 0$;
for drifting ones, $\Deltabar(x) = \sqrt{\Delta^2 - R^2(x)} \sim
\epsilon\sqrt{\delta - u \cos x}$, to leading order in $\epsilon$. Starting
from the origin of Figure~\ref{fig:u-delta} and moving counterclockwise
around the kidney bean, the corresponding graph of $\Deltabar(x)$ is zero for
the homoclinic locked state; flat for uniform drift states; modulated and
positive for modulated drift states; and partially zero/partially nonzero for
chimera states, with the fraction of drifting oscillators decreasing steadily
as we circulate back toward the origin.

Although we have focused on the chimera state in Eq.~\eqref{maineq}, it also
arises in other spatially extended systems. Indeed, it was first seen in
simulations of the complex Ginzburg-Landau equation with nonlocal coupling
\cite{kur02, kur03}.  That equation in turn can be derived from a wide class
of reaction-diffusion equations, under particular assumptions on the local
kinetics and diffusion strength that render the effective coupling nonlocal
\cite{kur02, kur03, shima04, tanaka}.

In two dimensions, the coexistence of locked and drifting oscillators
manifests itself as an unprecedented, bizarre kind of spiral wave---one
without a phase singularity at its center \cite{kur03, shima04}. Perhaps our
analysis of its one-dimensional counterpart can be extended to shed light on
this remarkable new mechanism of pattern formation.

\begin{acknowledgments}
Research supported in part by the National Science Foundation. We thank
Yoshiki Kuramoto for helpful correspondence, and Steve Vavasis for advice
about solving the self-consistency equation numerically.
\end{acknowledgments}



\end{document}